\documentclass[aps,pra,showpacs,twoside,twocolumn,10pt]{revtex4-1}
\usepackage[colorlinks=true, citecolor=blue, urlcolor=blue ]{hyperref}
\usepackage{epsfig,newlfont,amssymb,amsfonts,amsmath,bm,subfigure,palatino,mathtools,amsthm,braket,soul,enumitem,color,graphics,graphicx,times,physics,bbold,comment}
\usepackage[normalem]{ulem}

\newtheorem{proposition}{Proposition}

\begin{document}

\title{ Quantum battery with non-Hermitian charging}

\author{Tanoy Kanti Konar, Leela Ganesh Chandra Lakkaraju, Aditi Sen (De)}

\affiliation{Harish-Chandra Research Institute, A CI of Homi Bhabha National Institute,  Chhatnag Road, Jhunsi, Prayagraj - 211019, India
}

\begin{abstract}

We propose a design of a quantum battery  exploiting the non-Hermitian Hamiltonian as a charger.  In particular, starting with the ground or the thermal state of the interacting (non-interacting) Hamiltonian as the battery, the charging of the battery is performed via parity-time (\(\mathcal{PT}\))- and rotational-time (\(\mathcal{RT}\))-symmetric Hamiltonian to store energy. We report that such a quenching with a non-Hermitian Hamiltonian leads to an enhanced power output compared to a battery with a Hermitian  charger. We identify the  region in the parameter space which provides the gain in performance. We also demonstrate that the improvements persist with the increase of system size for batteries with both \(\mathcal{PT}\)- and \(\mathcal{RT}\)-symmetric chargers.  In the \(\mathcal{PT}\)-symmetric case, although the anisotropy of the $XY$ model does not help in the performance, we show that the $XXZ$ model as a battery with a non-Hermitian charger performs better than that of the $XX$ model having certain interaction strengths. We also exhibit that the advantage of non-Hermiticity remains valid even at finite temperatures in the initial states.


\end{abstract}

\maketitle

\section{Introduction}

Miniaturization of technology with the usage of quantum mechanical principles has become an intensive field of research in recent times. 
Notable achievements exhibiting quantum advantage over their classical analogs, thereby revolutionizing the arena of modern technologies include quantum key distribution \cite{gisinrmp}, quantum communication, \cite{tele1, dc},  quantum computers
\cite{nielsenchuang}, and devices for metrology \cite{Giovannetti2011} like quantum sensors \cite{degan2017} to name a few.  
In this respect, designing quantum thermal machines like the quantum refrigerator (QB) \cite{popescu10,chiara2020}, quantum battery \cite{Alicki, Batteryreview} and thermal transistor \cite{joulain16} has two motivations -- it is important to understand the concepts of heat, on one hand, work, and temperature in the microscopic limit, thereby developing the laws of quantum thermodynamics \cite{gemmer2004,qt2018,sai2016}  and on the other hand, how to achieve the optimal performance from the machines even when there is a competition between thermal and quantum fluctuations. It is also an interdisciplinary field lying at the crossroads of quantum optics, non-equilibrium statistical mechanics, and quantum information theory. Moreover, with the increase of control on quantum systems,  several experiments have been performed to verify quantum thermodynamical laws like the Jarzynski equality \cite{jarzynski1997,tiago2014,an2015} and thermal machines like quantum batteries  \cite{andolina2017,james2022} and quantum refrigerators \cite{nie2020}  in several physical substrates like  trapped ions, nuclear magnetic resonances, solid state systems, organic microcavity, and cold atoms etc. 



The original proposal for the QB considers the initial battery state as the ground state of a non-interacting Hamiltonian which can then be charged by global unitary operations \cite{Alicki, andolina2017,  campaioli2017, andolina2019, Batteryreview}. The main goal of such construction is to show that the work output or power stored (extracted) in (from) the battery gets enhanced 
in the presence of quantum mechanical systems or quantum operations. 
Instead of a non-interacting Hamiltonian, the ground or the thermal state of an interacting Hamiltonian can also be used as the battery \cite{Modispinchain,srijon2020} while the local magnetic field in a suitable direction is applied at each site to maximize the energy storage of the battery. Such a design turns out to be appropriate even in the presence of decoherence and disorder \cite{Giovannetti2019, srijon2020, srojon20} as well as in higher dimensions \cite{srijon2021}.

The evolution of a quantum system is described by a Hamiltonian which is typically a Hermitian operator. It was shown that, relaxing the Hermiticity condition, if one considers non-Hermitian systems with parity-time (\(\mathcal{PT}\)) symmetry \cite{Bender'98,Bender2007} (with \(\mathcal{P}\) being the reflection operator in space and \(\mathcal{T}\) being the time-reversal operator)  or rotation-time  (\(\mathcal{RT}\)) symmetry (with \(\mathcal{R}\) being the rotation operator along a fixed axis) \cite{SongReal}, the energy eigenvalues can be real depending on the system parameters, thereby maintaining all the properties of standard quantum mechanics and describing natural processes. However, such a system undergoes a transition from a broken to an unbroken phase where the energy spectrum becomes real from imaginary values, known as exceptional points \cite{Bender'98, Bender2007}. Several counter-intuitive results are also reported in this framework -- when a local \(\mathcal{PT}\)-symmetric Hamiltonian acts on a part of an entangled state, it was shown that there is a violation of the no-signaling principle \cite{lee2014} which was later settled by Naimark's dilation \cite{Tang2016}. On the other hand, interesting phases in the ground state of the  \(\mathcal{RT}\)-symmetric Hamiltonian are also reported \cite{SongReal, PTSymmetry_spin1, PTSymmetry_spin2} in which the broken-unbroken transition is found to be connected with the factorization surface of the corresponding Hermitian models \cite{ganesh2021}.  
Over time, it has been realized that such systems can have great influences in different branches of physics ranging from optics \cite{guo2009,Wimmer2015} to electronics \cite{schindler2011}, Bose-Einstein condensates \cite{kreibich2014} and many-body physics \cite{SongReal, PTSymmetry_network,PTSymmetry_tighbinding_2, PTSymmetry_tighbinding_4, PTSymmetry_tightbinding_1, PTSymmetry_spin1, PTSymmetry_spin2}. Further, it was shown that the performance of quantum sensors can also be improved with non-Hermiticity \cite{ptsensing1,ptsensing2, nonhermitian_top_sensor1, nonhermitian_top_sensor2, pseduo_herm_sensor,mcdonald2020}. Interestingly, the optimal time required to evolve the initial state to an orthogonal one can be made arbitrarily smaller with the non-Hermitian Hamiltonian \cite{wang_cpb_2020} while the time scaling technique can also enhance the quantum control protocol in non-Hermitian systems \cite{impens_pra_2021}.

Motivated by the advantages provided by non-Hermitian systems, we utilize non-Hermiticity to propose a set-up of a quantum battery. In particular, when the initial state of the battery is the ground states of the interacting and non-interacting Hamiltonian, we use  \(\mathcal{PT}\)- as well as \(\mathcal{RT}\)-symmetric Hamiltonian to charge the battery. In both cases, we show that the power of the battery gets enhanced with the help of non-Hermitian charging Hamiltonian compared to their Hermitian counterparts. In particular, we identify a parameter region where such a beneficial role can be found. We demonstrate that the maximum power decreases with the anisotropy parameter of the $XY$ model as a battery in the \(\mathcal{PT}\)-symmetric case and  as a charger in the \(\mathcal{RT}\)-symmetric scenarios although, for a fixed anisotropy, non-Hermiticity still provides a benefit over the Hermitian set-up. We also observe that the  energy that can be extracted, measured  via ergotropy \cite{Alicki}, coincides with the work output in the evolution and hence the power computed  here quantifies both the storage as well as extractable power of the QB.

Moreover, the trend of the maximum power saturates to a finite value for a moderate system size for batteries with both \(\mathcal{PT}\)- and \(\mathcal{RT}\)-symmetric chargers. 
When the initial state is the thermal state of the system, the maximum power decreases with the increase of temperature although some distinct behavior due to non-Hermitian evolution is observed in the limit of infinite temperature.

The implementation of the non-Hermitian quantum battery proposed here, especially, the $\mathcal{PT}$-symmetric charging can be carried out by embedding the charging Hamiltonian in a higher dimensional space with suitable modifications. The original idea \cite{OriginalNaimark} suggests that the higher dimensional system whose subsystem's dynamics is governed by the \(\mathcal{PT}\)-symmetric Hamiltonian evolves under a unitary operation, thereby raising a possibility to simulate in experiments 
\cite{Tang2016}. The caveat is that, this way of simulating the $\mathcal{PT}$-symmetric Hamiltonian should be in the unbroken region. It was recently shown that the efficiency in measuring a parameter can be enhanced around the exceptional point of a $\mathcal{PT}$-symmetric Hamiltonian \cite{weakexperiment}.  A very similar idea of embedding the Hamiltonian in a higher-dimensional system was also proposed to design a non-Hermitian sensor where weak measurements are performed in order to get the results in the broken region \cite{ptsensing1} (see also \cite{weakmeasure2}). All these proposals as well as implementations indicate that the  advantage at the unbroken-broken transition point, i.e., at the exceptional point reported here
can also be obtained while realizing quantum batteries in a non-hermitian domain.  In our work, we also show that the non-Hermitian charger can be realized as an effective Hamiltonian of a quantum system interacting with the environment. Specifically, if the quantum jumps are ignored in the master equation, the state undergoes an evolution corresponding to a non-Hermitian Hamiltonian or a charger in our case.

The paper is organized in the following manner. In Sec. \ref{sec:figureofmerit}, we set the stage by introducing the quantities that quantify the performance of the battery. The design of the battery and its performance when it is charged with the \(\mathcal{PT}\)-symmetric Hamiltonian are presented in Sec. \ref{sec:PTsymmetric} The results obtained when the charger has  \(\mathcal{RT}\) symmetry are discussed in Sec. \ref{sec:RTsymm}. The concluding remarks are given in the last section, Sec. \ref{sec:conclu}.

\section{Modelling quantum Battery and its figures of merits }
\label{sec:figureofmerit}

The design of a quantum battery has two components -- 1. the battery Hamiltonian, and 2. a charger. In this paper, we choose both ground and the thermal states of interacting as well as non-interacting Hamiltonians, \(H_B\), as the initial state of the battery. The details of these Hamiltonians will be discussed in succeeding sections. 

\subsection{Non-Hermitian realization of charger in open-system framework}
In general, a charging Hamiltonian is used to excite the particles to a higher energy state so that a high amount of energy gets stored in the QB which can be extracted from the battery in a suitable time by a unitary operation. In this paper,  instead of a Hermitian Hamiltonian, two non-Hermitian Hamiltonians having parity-time (rotation-time) symmetry, \(H_{charging}^{\mathcal{PT} (\mathcal{RT})}\) are used independently as chargers of the battery. Specifically, we use the well-known quantum $\mathcal{PT}$-symmetric Hamiltonian \cite{lee2014} and $\mathcal{RT}$-symmetric $XY$-model \cite{SongReal} for the purpose of charging (for details, see Secs. \ref{sec:PTsymmetric} and \ref{sec:RTsymm}).  Let us now describe how the non-Hermitian evolution can be observed in the dynamics of open quantum systems. When a system Hamiltonian, denoted as \(H_S\), is coupled to an external environment with a decay rate \(\gamma\), the dynamics is described by the Gorini-Kossakowski-Lindblad-Sudarshan (GKLS) master equation  \cite{breuer2002} as
 \begin{eqnarray}
     \nonumber\frac{d\rho}{dt}&=&-i[H_S,\rho]+\gamma[\mathcal{L}\rho \mathcal{L}^\dagger-\frac{1}{2}\{\mathcal{L}^\dagger \mathcal{L},\rho\}]\\&=&-i(H_{eff}\rho-\rho H_{eff}^\dagger)+\gamma\mathcal{L}\rho \mathcal{L}^\dagger,
 \end{eqnarray}
 where \(H_{eff}=(H_S-i\frac{\gamma}{2}\mathcal{L}^\dagger \mathcal{L})\) is  the effective Hamiltonian which is non-Hermitian
 due to the dissipative processes. It incorporates the original system Hamiltonian while accounting for the decay induced by the environment. Additionally, the term \(\frac{1}{2}\mathcal{L}\rho \mathcal{L}^\dagger\) in the GKLS master equation is referred to as the jump operation. It represents the contribution to the system dynamics caused by quantum jumps, which occur due to the environment. However, in the semi-classical limit, it is often permissible to neglect the effects of the jump operation. Consequently, the evolution of the system is governed by the non-Hermitian effective Hamiltonian \(H_{eff}\) \cite{roccati_osi_2022}. Within the semi-classical approximation, we can focus 
 on the non-Hermitian evolution of the system, which is given as
  \begin{eqnarray}
     \frac{d\rho}{dt}=-i(H_{eff}\rho-\rho H_{eff}^\dagger);\quad \rho(t)=\frac{\rho(t)}{\text{Tr} [\rho(t)]}.
     \label{eq:evolution}
 \end{eqnarray}
 More specifically, in our case, the evolution is governed by 
 \begin{equation}
     \frac{d\rho}{dt}=-i[H_{charging}^{\mathcal{PT} (\mathcal{RT})}\rho-\rho (H_{charging}^{\mathcal{PT} (\mathcal{RT})})^\dagger], 
 \end{equation}
 such that the dynamical state \(\rho(t)\) is obtained after evolving the system with a non-Hermitian Hamiltonian as \(\rho(t) = (1/\mathcal{N})\exp(-i H_{charging}^{\mathcal{PT} (\mathcal{RT})}t) \rho(0) 
 \exp (i H_{charging}^{\mathcal{PT} (\mathcal{RT})}t)\) with \(\mathcal{N} = \mbox{Tr}[\exp(-i H_{charging}^{\mathcal{PT} (\mathcal{RT})}t) \rho(0) 
 \exp (i H_{charging}^{\mathcal{PT} (\mathcal{RT})}t)]\). Notice that unlike unitary dynamics governed by a Hermitian Hamiltonian, we need to normalize the evolved state at each time interval in the non-Hermitian domain.  
 In the succeeding sections,  we will explicitly discuss the specific form of the system and bath Hamiltonian from which the effective non-Hermitian Hamiltonian as charger considered in this paper emerges.

The performance of a quantum battery is decided by the amount of generated power.  In order to describe that, we need the thermodynamic definition of work.
\subsection{Performance quantifiers: Work and power} The work output at a given time instance can be measured as \cite{Alicki, Batteryreview} 
 \( W(t) = \tr[H_B(\rho(t) - \rho(0))]\), 
 where \(\rho(0)\) is the initial state of the battery Hamiltonian, which are taken to be the ground or the thermal states of \(H_B\). 
The maximal power can be computed by performing maximization over time as
\begin{equation}
   P_{max}  = \max_t \frac{W(t)}{t} = \max_t P(t),
   \label{Eq:maxP}
\end{equation}
where \(P(t)\) denotes the average power at some time, \(t>0\). In our case, even in the presence of non-Hermiticity, \(P(t)\) is always found to be real. In our paper, our primary objective is to design a battery that maximizes energy storage capacity by using a non-Hermitian charger.  To achieve maximum energy storage, we commence the charging process from the ground state of the Hamiltonian. This choice is motivated by the fact that the ground state exhibits the maximum separation from the highest excited state.
However, when it comes to the maximum power in terms of charging,  it also optimizes the time required to reach the highest excited state and hence the excited state may provide a higher \(P_{\max}\) than that of the ground state. 

 In general, when the value of a parameter, e.g., the applied magnetic field, increases,    the amount of power generated trivially increases. In order to maintain a fair comparison between different situations,  we normalize the battery Hamiltonian as
 \begin{eqnarray}
\frac{1}{E_{\max}-E_{\min}}[2 H_{B} - (E_{\max}+E_{\min}) \mathrm{I}] \rightarrow H_{B},  
\end{eqnarray}
where the minimum and maximum eigenenergies are denoted by $E_{\min}$ and $E_{\max}$ respectively. Thus, the spectrum is bounded between $[-1,1]$ which ensures that the advantage is not an artifact of the parameters. 

As mentioned, the energy stored in the battery can be represented as \(W(t)\) although the entire energy may not be extractable. In other words, the energy that can be extracted from the battery may not always coincide with the work output in several scenarios including when the battery is in contact with the environment \cite{Giovannetti2019, srojon20}.  The extractable energy, known as  ergotropy, from the battery at some time instance \(t\) can be quantified as \cite{Alicki, Giovannetti2019, Munro20}

\begin{eqnarray}
\mathcal{E} =  E_B(t) - \min_{U}  \tr(H_B U \rho(0) U^\dagger), 
\end{eqnarray}
where $E_B(t) = \tr(H_B \rho(t))$ is the average energy at some time instant and the minimization is over all possible charging unitary operators. We will later show that the ergotropy is the same as the energy stored in the situation considered here.  





\section{Enhancement of power with $\mathcal{PT}$ symmetric charger }
\label{sec:PTsymmetric}

Let us describe briefly the set-up of a quantum battery and the charger in the non-Hermitian framework. 
The ground or the thermal state of the $XYZ$ Hamiltonian in the presence of an external magnetic field, given by
    \begin{eqnarray}
        H_{B} &=&\frac{J}{4}\sum_{r=1}^{N}\left[(1+\gamma)\sigma^{x}_{r}\sigma^{x}_{r+1}+(1-\gamma)\sigma^{y}_{r}\sigma^{y}_{r+1}\right]\nonumber\\&&+\frac{\Delta}{4}\sum_{r=1}^N\sigma_r^z\sigma_{r+1}^z+\frac{h}{2}\sum_{r=1}^{N}\sigma_r^z,
        \label{eq:batteryint}
    \end{eqnarray}
    act as the battery. Here $\sigma^i$, $i \in \{x,y,z\}$ matrices represent the Pauli matrices, $\gamma$ corresponds to the anisotropy in the $xy$-plane, $J$ and $\Delta$ are the coupling constants in the \(xy\)-plane and \(z\) direction respectively and \(h\) is the strength of the magnetic field.  We consider the open-boundary condition. Notice that with available technologies, the above Hamiltonian can be controlled and manipulated using physical systems like cold atoms, trapped ions, and nuclear magnetic resonances \cite{bloch03, iontrapBlatt, aditireview, nmr_aditi}.

\subsection{\(\mathcal{PT}\)-symmetric Hamiltonian as a Charger}

The quantum battery is charged by using a local $\mathcal{PT}$-symmetric Hamiltonian which can be simulated in the laboratory \cite{Tang2016,Li2019} as a dilation of higher dimensional Hilbert space \cite{unther2008,OriginalNaimark,huang2019}. It is expressed as
\begin{equation}
    H_{charging}^\mathcal{PT}=\sum_{j=1}^{N}\sigma_j^x+i\sin \alpha_i \sigma_j^z, 
    \label{eq:chargingPT}
\end{equation}
where the Hamiltonian possesses parity symmetry, i.e., \(\mathcal{P}\) acts on the Hamiltonian, \(\mathcal{P}  H_{charging}^\mathcal{PT} \mathcal{P} = H_{charging}^{*\mathcal{PT}}\) where \(\mathcal{P}=\sigma^x\) while  $\mathcal{T}$ is a complex conjugation in finite dimension, $\mathcal{T}i\mathcal{T}^{-1} = -i$, where $i = \sqrt{-1}$.
 Here \(\alpha_i\) is the $\mathcal{PT}$-symmetry (non-Hermiticity) parameter of  \(H_{charging}^\mathcal{PT}\)  and $\alpha_i=\pi/2$ represents the exceptional point where eigenvectors and eigenvalues of the local charging Hamiltonian coalesce. 
 
Let us now describe how the charging Hamiltonian in Eq. (\ref{eq:chargingPT}) can be obtained from the open quantum system formalism. We first consider a single spin interacting with a continuum mode of the electromagnetic field acting as a bath where the spontaneous emission takes place. The Hamiltonian of the single spin is represented as \(H_S=\sigma^x_S\) and the evolution of the system is governed by the GKLS master equation \cite{roccati_osi_2022}:
\begin{equation}
    \frac{d\rho}{dt}=-i[\sigma^x_S ,\rho]+\gamma (\sigma^-\rho\sigma^+-\frac{1}{2}\{\sigma^+\sigma^-,\rho\}),
\end{equation}
for which the effective Hamiltonian can be written as
\begin{eqnarray}
     \nonumber H_{eff} & =&\sigma^x-\frac{i\gamma}{2}\sigma^+\sigma^-\\&=&\sigma^x-\frac{i\gamma}{4}\sigma^z.
     \label{eq:HeffPT}
 \end{eqnarray}
Here $\sigma^\pm = \sigma^x \pm i \sigma^y$ are the Lindblad operators due to the environment where the constant imaginary shift is ignored \cite{imaginaryfield_ueda_2018, hiddentransition_2022}. In our calculations, instead of using the dissipation term \( \frac{\gamma}{4}\), we choose the non-Hermiticity parameter \(\sin\alpha_i\) \cite{keck_pra_2018}. At \(\alpha_i=0\), the Hamiltonian reduces to the Hermitian one. 
Instead of taking \(\alpha_i=0\) which changes  the magnetic field only in the \(z\) direction, we  can consider Hermitian charger as 
\begin{equation}
H_{charging}^{Herm}=\sum_{j=1}^{N}\sigma_j^x+ \sin \alpha_r \sigma_j^z,
    \label{eq:charginghermi}
\end{equation}
where $\alpha_r$ parameterizes the Hermitian Hamiltonian and the corresponding average power of the QB is denoted by \(P^{Herm}(t, \alpha_r)\) while \(P^{\mathcal{PT}}(t, \alpha_i)\) represents the same with the \(\mathcal{PT}\)-symmetric charger (see Appendix \ref{sec:appendix}).  
Before proceeding to compute the maximal power produced from the battery, let us first establish a relation between the ergotropy and the work output.

\begin{proposition}
    The ergotropy or the extractable work coincides with the energy stored in the system when the initial state is the ground state of the battery Hamiltonian and it does not depend on whether the evolution is governed by the Hermitian or non-Hermitian Hamiltonian.
\end{proposition}

\begin{proof}
    
 The ergotropy (\(\mathcal{E}\)) is defined as \(\mathcal{E}=\text{Tr}(\rho(t)H_B)-\underset{U}{\min}(U\rho U^{\dagger} H_B)\), which is the maximum amount of energy extractable from the system. If the initial state of the system is written in the spectral decomposition as $\rho = \sum_{k=1}^{N} r_{k} \ket{r_k}\bra{r_k}$, where $r_k$s are arranged in the ascending order, it has been shown  \cite{allahverdyan2004maximal} that the minimization in the second term of $\mathcal{E}$ occurs for a specific state called the \textit{passive state} (\(\rho_p\)) associated with the Hamiltonian \(H_B\). The passive state in the energy eigenbasis can be expressed as
\begin{equation}
    \rho_p=\sum_{k=1}^{N} r_{N-k}\ket{\epsilon_k}\bra{\epsilon_k},
\end{equation}
where \(r_k\)s represent the eigenvalues of the state and \(\ket{\epsilon_i}\)s denote the eigenspectrum of the battery Hamiltonian which are arranged in the ascending order.  Therefore, ergotropy expression becomes $\mathcal{E} = \text{Tr}(\rho(t)H_B) - \text{Tr}(\rho_pH_B) = E_B(t) - \sum_k r_{N-k} E_k$, where $E_k$ are the ordered eigenenergy of $H$. Note that if the initial state is pure, the final state after the unitary evolution also remains pure. Suppose, the decomposition contains only a single term, i.e., $r_1 = 1$ and $\forall ~k \ne 1, ~ r_k = 0 $, \(\rho_p=\ket{\epsilon_1}\bra{\epsilon_1}\), representing the ground state of the battery Hamiltonian. Thus, if the initial state is in the ground state, the corresponding energy stored is $W(t) = \text{Tr}(\rho(t)H_B)-\text{Tr}(\rho(0)H_B)$ which coincides with \(\mathcal{E}\). 
Let us now argue that the above proposition also holds for a non-Hermitian charger. The non-Hermitian system can be described as an effective Hamiltonian of a Hermitian system,  interacting with the bath, without the jump operators, as shown in Eq. (\ref{eq:HeffPT}).  The evolution can then be seen as $\ket{\psi(t)} = e^{-iH_{eff} t} \ket{\psi(0)}$, where $H_{eff}$ is a non-Hermitian Hamiltonian \cite{franco_2019}. Thus in the semi-classical limit, if the initial state is pure, it remains so, even after undergoing evolution corresponding to a non-Hermitian Hamiltonian \cite{barch2023scrambling}. Thus, if the initial state is the ground state which is the case considered in this paper, the above proof holds for the non-hermitian Hamiltonian for all values of \(\alpha_i\). 
\end{proof}


\begin{figure}[h]
    \centering
    \includegraphics[scale=0.6]{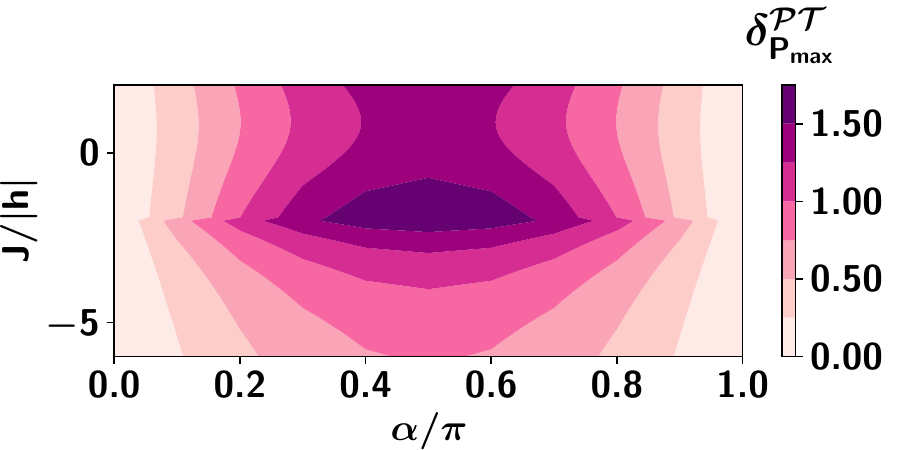}
    \caption{(Color online.) \textbf{Advantage of non-Hermitian charging over Hermitian ones in terms of the maximum power}.   Map plot of $\delta_{P_{\max}}^{\mathcal{PT}}$, defined in Eq. (11),  with respect to the parameters, \(\alpha/\pi\) (\(x\)-axis) and \(J/|h|\) (\(y\)-axis). In the entire parameter regime considered here,  a non-vanishing advantage is obtained, thereby establishing the benefit of a non-Hermitian $\mathcal{PT}$-charger  in Eq. (\ref{eq:chargingPT}) over the Hermitian one (Eq. (\ref{eq:charginghermi})). The initial state is the ground state of the battery Hamiltonian.  Note that for the non-Hermitian charger, \(\alpha = \alpha_i\) while  \(\alpha = \alpha_r\) for the Hermitian one.    All the axes are dimensionless. }
    \label{fig:power_nh_h}
\end{figure}

Before considering the general battery Hamiltonian, let us first illustrate the effects of non-Hermiticity on the performance of the QB when the initial state is the ground state of the \(XX\) model, i.e., \(H_B\) with \(\gamma =0\) and \(\Delta =0\). Let us now analyze  \(P_{\max}\) obtained from the \(\mathcal{PT}\)-symmetric and Hermitian chargers.   In these scenarios, when there are two sites, we obtain the following proposition on enhancement due to non-hermiticity. 

\begin{proposition}
    The maximum power output of the battery made out of two lattice sites in the presence of the $\mathcal{PT}$-symmetric charger,  \(P^{\mathcal{PT}}_{\max}\), is higher than that of a QB which is charged by the Hermitian Hamiltonian when the initial state is the ground state of the \(XX\) model.
\end{proposition}


\begin{proof} The ground state, $\ket{\psi(0)}$ 
as the initial state of the $XX$ model takes the form
$\ket{0001}$ in the computational basis
when $J \in \{ -2h, 2h-0.1 \}$. 
After evolution with the local $\mathcal{PT}$-symmetric charging Hamiltonian, the evolved state at time $t$, \(|\psi(t)\rangle\), can be expressed (see Appendix) as a function of the non-Hermitian parameter, \(\alpha_i\), and system parameters, \(J\), \(h\) and time \(t\). We can then straightforwardly compute the maximal power both for Hermitian and non-Hermitian cases (see Appendix for the expressions). 
To prove the enhancement due to the non-Hermitian charger, we consider the quantity, called the difference in maximum power between non-Hermitian and Hermitian domains, given by
\begin{eqnarray}
 \delta_{P_{\max}}^{\mathcal{PT}} =   \nonumber \max_t(P^\mathcal{PT}(t))-\max_t(P^{\text{Herm}}(t)) = P_{\max}^\mathcal{PT}-P_{\max}^{\text{Herm}}, \\&&  
 \label{eq:diffPT}   
\end{eqnarray}
which is also a function of \(\alpha_i\) and \(\alpha_r\).
 We  provide the analytical form of the average power $P(t)$ for non-Hermitian as well as Hermitian scenarios with respect to the other parameters of the system in  Appendix \ref{sec:appendix} (Eqs. (\ref{eq:power_nherm}) and (\ref{eq:power_herm})). However, as it is apparent from those expressions, maximizing with respect to time and finding the difference between powers with non-Hermitian and Hermitian chargers would be cumbersome and it cannot take a compact form. 
Instead, we plot the difference between the maximum power generated via non-Hermitian and Hermitian chargers, $\delta^{\mathcal{PT}}_{P_{max}}$,  in Fig. \ref{fig:power_nh_h}. For the entire region of the interaction strength $J/|h|$ and the range of interactions, \(\alpha_{i} (\alpha_r) \in [0,\pi]\), we find that $\delta^{\mathcal{PT}}_{P_{max}}>0$, thereby demonstrating the advantage of non-Hermiticity over the Hermitian Hamiltonian.

\end{proof}



\begin{figure}
    \centering
    \includegraphics[scale=0.5]{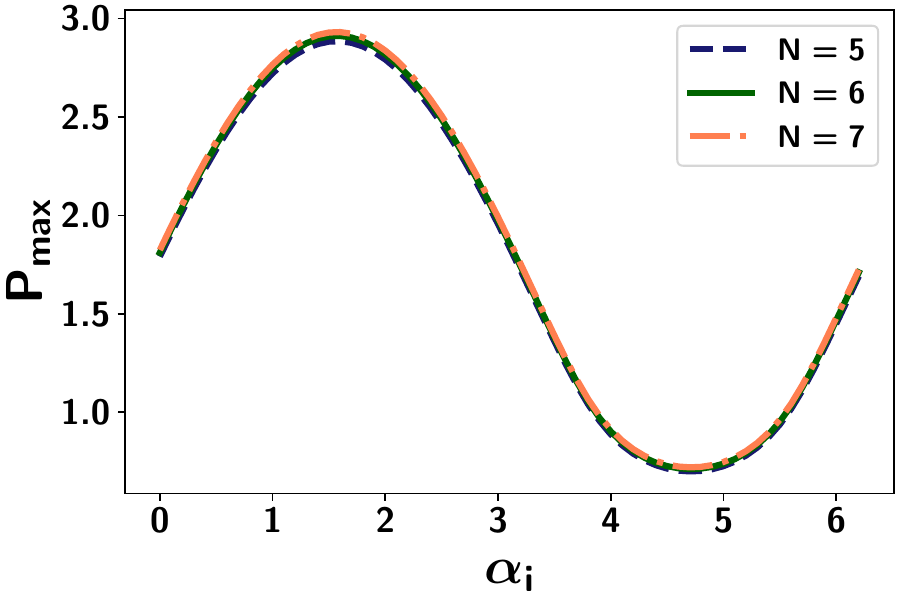}
    \caption{(Color online) {\bf Role of non-Hermiticity.} Maximum power output, \(P_{\max}\) vs. non-Hermiticity parameter \(\alpha_i\) in the charger. The effect of an increase in system size is also depicted by taking different \(N\) values. The initial state of the QB is again taken to be the ground state of the battery. 
    Here \(J/|h|= 1\) and \(\Delta =0\) in the battery Hamiltonian, \(H_B\) in Eq. (\ref{eq:batteryint}). 
    All the axes are dimensionless.}
    \label{fig:inspower_alpha}
\end{figure}


Let us now illustrate that the advantage persists even with the increase of system sizes, in presence of anisotropy in the battery Hamiltonian and exchange interaction in the \(z\) direction, i.e., with the \(XXZ\) model. The Proposition 2 shows that for a given \(\alpha_i\), \(\delta_{P_{\max}}^{\mathcal{PT}}\) is nonvanishing. 

\begin{figure}
    \centering
    \includegraphics[scale=0.5]{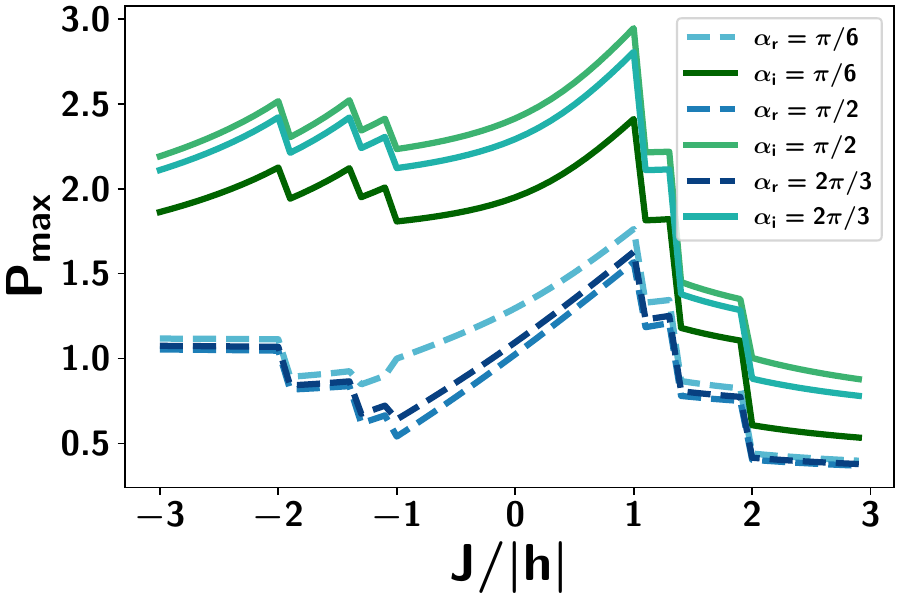}
    \caption{(Color online) \textbf{Interaction dependence.} \(P_{\max}\) (ordinate) against $J/\abs{h}$ (abscissa).  Different curves represent different non-Hermiticity and Hermiticity parameters, \(\alpha_i\), and \(\alpha_r\)  in Eqs. (\ref{eq:chargingPT}) and (\ref{eq:charginghermi}) respectively.  The \(\mathcal{PT}\)-symmetric local charger, given in  Eq. (\ref{eq:chargingPT}), is applied at each site of the battery in the non-Hermitian case while the Hermitian charging Hamiltonian in Eq. (\ref{eq:charginghermi}) is used in the Hermitian scenario. 
    In both situations, the $XX$ model acts as the QB. 
    Solid lines represent \(P_{\max}\) via non-Hermitian chargers, having higher power than that of the Hermitian ones (the dashed lines).
    Here $N=8$. All the axes are dimensionless.}
    \label{fig:inspower_J}
\end{figure}

\subsection{Effects of non-Hermiticity and interactions on the QB} We first examine the pattern of maximal extractable power \(P_{\max}\) from the QB with the variation of \(\alpha_i\) and the interaction strength in the \(xy\) plane. A few observations immediately emerge from Figs. \ref{fig:inspower_alpha} and \ref{fig:inspower_J}. Since the charging Hamiltonian involves \(\sin \alpha_i\), the maximal power also shows a periodic nature with \(\alpha_i\) as depicted in Fig. \ref{fig:inspower_alpha}. In order to compare the power generated by the charging Hamiltonian possessing \(\mathcal{PT}\)-symmetry in Eq. (\ref{eq:chargingPT}) with \(\alpha_i \neq 0\) and by the Hermitian charger, given in Eq. (\ref{eq:charginghermi}),  we find that 
    \begin{equation}
        P_{\max}^{Herm}  < P_{\max}^{\mathcal{PT}} \, \, \mbox{for} \, \,  (0 <\alpha_i, \alpha_r < \pi),
    \end{equation}
and we drop the superscripts $\mathcal{PT}$ and ``Herm" in the analysis as it will be
evident from the context. Note that in this region,  the charging via \(\mathcal{PT}\)-symmetric Hamiltonian has a real energy spectrum \cite{lee2014}. \(P_{\max}\) reaches its maximum value with the charging Hamiltonian having \(\alpha_i =\pi/2\) which is the exceptional point, thereby establishing the dependency of power on non-Hermiticity. 
Furthermore, we notice that the optimal state obtained during the maximization of time required to reach the orthogonal state from the initial one \cite{wang_cpb_2020, impens_pra_2021} may not always be the excited state from which \(P_{\max}\) is achieved.  To achieve the maximum power output, the energy stored (i.e., the numerator of \(P_{\max}\)) is higher in the non-Hermitian case than that of the Hermitian one while the time in  \(P_{\max}\) may not be smaller in the former case than in the latter one.

The performance of the battery remains almost invariant with the increase of system size \(N\) of the QB Hamiltonian although the scaling analysis of the QB requires much more careful investigation which we will do next. It was shown that the QB can show quantum advantage (i.e., a battery is said to provide a quantum advantage when \(P_{\max}\) is higher for the battery Hamiltonian having non-vanishing interaction strength \(J/|h| \neq 0\) than that of the battery with vanishing interaction strength, i.e.,  \(J/|h|=0\)) when the initial state of the battery is the ground state of the $XX$ model \cite{srijon2020}. We observe in Fig. \ref{fig:inspower_J} that the non-Hermitian charging Hamiltonian can also furnish quantum advantage for different values of \(\alpha_i\).  Moreover, unlike the Hermitian charger, the quantum advantage in the non-Hermitian framework is observed both in the positive and  negative regions of \(J/|h|\) although the sharp continuous increase in the positive domain is not visible in the negative domain. 

\begin{figure}
    \centering
    \includegraphics[scale=0.5]{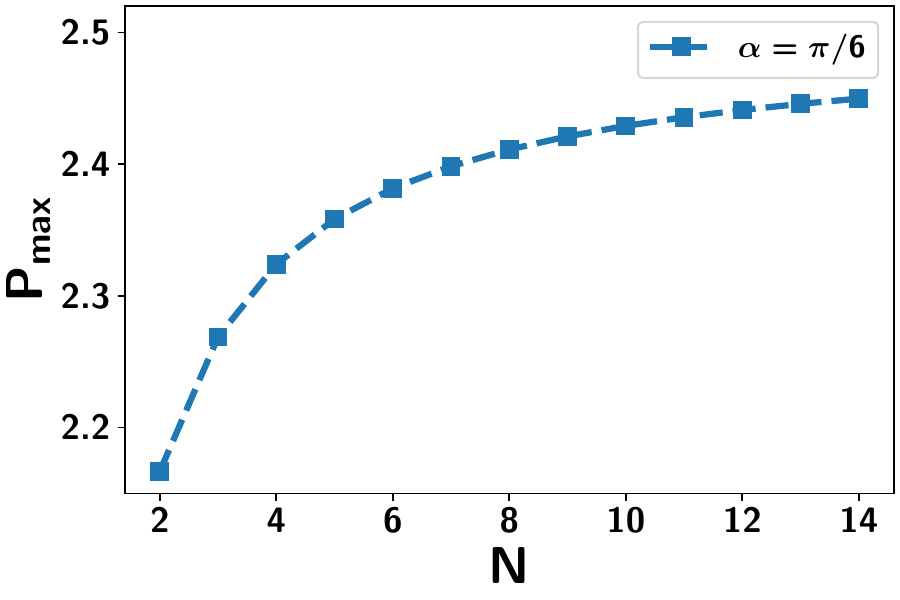}
    \caption{(Color online) \textbf{Scaling.}  \(P_{\max}\) (ordinate) as a function of $N$ (abscissa). All other specifications are the same as in Fig. \ref{fig:inspower_alpha}. Fitting the data shows that \(P_{\max} \propto \sqrt{N}\). 
Both the axes are dimensionless.}
    \label{fig:power_N}
\end{figure}

\subsection{ Scaling analysis of QB} We now explore the quantum advantage in our model with the increase of sites in the lattice. For a fixed \(\alpha_i >0\), we find that \(P_{\max}\) increases monotonically with \(N\) as depicted in Fig. \ref{fig:power_N}. 
More careful analysis reveals that  $P_{max}$ scales not linearly with the system size. Specifically, we find
$$P_{\max} \propto\sqrt{N},$$ 
when the initial state of the QB is the ground state of the $XX$ model with open boundary condition and the maximum power saturates slowly with system size. Upon considering a periodic boundary condition, however, we find that no such scaling occurs, i.e.,  we find that \(P_{\max}\) remains constant as the system size increases.

\begin{figure}
    \centering
    \includegraphics[scale=0.5]{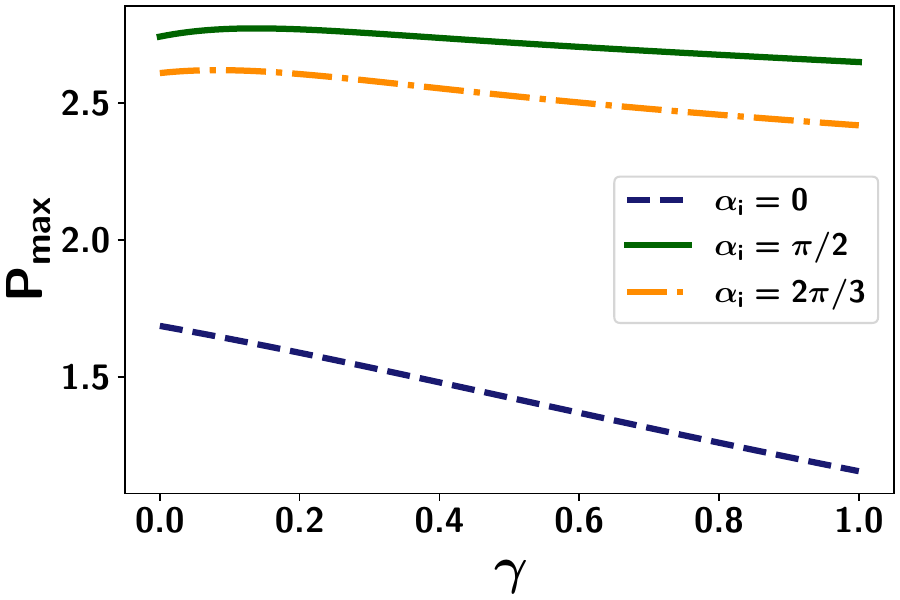}
    \caption{(Color online) \textbf{ Dependence on anisotropy.}
    \(P_{\max}\) (vertical axis) with  $\gamma$ (horizontal axis) of the QB Hamiltonian for different values of \(\alpha_i\). \(\alpha_i =0\) represents the battery with a  Hermitian charger. Other specifications are the same as in Fig. \ref{fig:inspower_alpha}. 
    All the axes are dimensionless.}
    \label{fig:power_gamma}
\end{figure}

\begin{figure}
    \centering
    \includegraphics[scale=0.5]{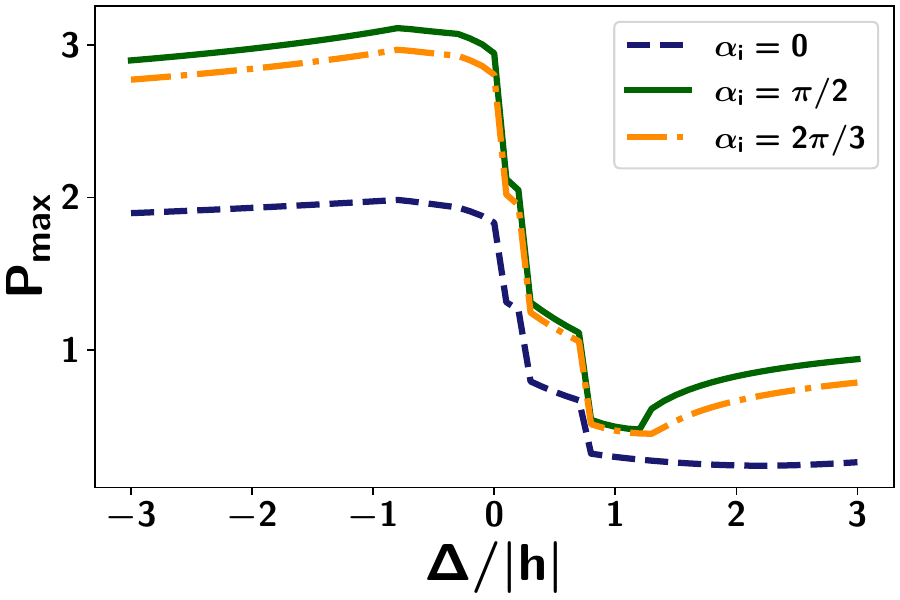}
    \caption{(Color online) \textbf{Importance of the $XXZ$ model as battery.} Trends of maximum power output, \(P_{\max}\) with respect to the variation of interaction strength in the \(z\) direction, \(\Delta/|h|\) for different values of \(\alpha_i\). Clearly, we observe that the \(XXZ\) model as a battery has some beneficial role over the $XX$ model with \(J/|h|=1.0\). 
 All other specifications are the same as in Fig. \ref{fig:inspower_alpha}. 
 Both the axes are dimensionless.}
    \label{fig:power_pt_ZZ}
\end{figure}

\subsection{Role of anisotropy and coupling in the \(z\)-direction} Upto now, the entire analysis is carried out when the initial battery Hamiltonian is the $XX$ model. As shown in the Hermitian case \cite{srijon2020,srijon2021}, the presence of anisotropy in the QB Hamiltonian typically suppresses the performance, i.e., \(P_{\max}\) decreases with \(\gamma\) for a fixed \(\alpha_i\) and \(J/|h|\) which are chosen in the region where quantum advantage is seen (see Fig. \ref{fig:power_gamma}). However, for nonvanishing \(\alpha_i\), we find that the rate of decrease in \(P_{\max}\) after a certain anisotropy parameter diminishes with \(\gamma\), i.e., after a decrease with \(\gamma\), \(P_{\max}\) almost saturates for \(\gamma > 0.5\) which was absent in the Hermitian counterpart as shown with \(\alpha_i =0\).    

The introduction of interaction in the \(z\) direction also leads to a non trivial effect on QB's power extraction -- for a fixed \(J/|h|\), we find that with the decrease of \(\Delta/|h| (<0)\), \(P_{\max}\) increases for different values of \(\alpha_i\) and the maximum \(P_{\max}\) is again obtained with the symmetry breaking transition point, i.e., \(\alpha_i = \pi/2\) (as shown in Fig. \ref{fig:power_pt_ZZ}).


\begin{figure}
    \centering
    \includegraphics[scale=0.5]{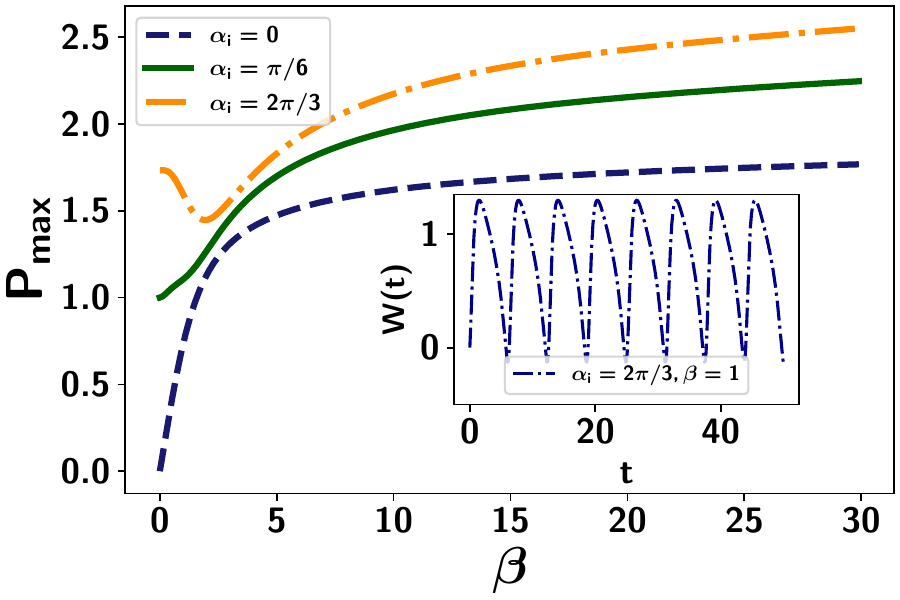}
    \caption{(Color online) \textbf{Temperature-dependence of the initial state: }
   \(P_{\max}\) (ordinate) vs. $\beta = 1/k_BT$ (abscissa) for different values of \(\alpha_i\).
   The initial state of the QB is prepared as the canonical equilibrium state of the $XX$ model while the charger is the \(\mathcal{PT}\)-symmetric one with \(\alpha_i \neq 0\). Notice that the decreasing behavior of the power with the increase of temperature is the same as typically observed in the Hermitian domain. However, close to high temperature, in the framework of non-Hermitian systems, we find some different behavior than the one in the Hermitian paradigm. 
   (Inset) \(W(t)\) with \(t\) for a fixed \(\beta =1.0\). It shows that \(W(t)\) goes negative which is responsible for a non-monotonic behavior of \(P_{\max}\) at high temperature.
   All the axes are dimensionless.}
    \label{fig:power_thermal}
\end{figure}

\subsection{Thermal state as initial state of QB}

It is not possible to achieve the exact ground state of any Hamiltonian in laboratories. In particular,  noise due to thermal fluctuation is unavoidable. 
To incorporate this imperfect situation, let us take the thermal state of the form, $\frac{\exp(-\beta' H_B)}{\mbox{Tr}{\exp(-\beta' H_B)}}$ where \(\beta' = 1/k_B T\) is inverse temperature (\(k_B\) being the Boltzmann constant and \(T\) being the temperature and we take \(\beta = \beta'/|J|\)) as the initial state of the QB. First of all, as one expects, we obtain the maximum power output from the battery even in the non-Hermitian framework, when the temperature of the thermal state is moderately low and \(P_{\max}\) monotonically decreases with the increase of temperature (the decrease of \(\beta\)). 

At high temperatures, a certain abnormality arises in the non-Hermitian regime.
In this respect, notice that with \(\beta \rightarrow 0\), i.e., in presence of infinite temperature, the thermal state of a Hermitian Hamiltonian, \(H_B\), reduces to the maximally mixed state. 
When the charging Hamiltonian is Hermitian and when the initial state is a thermal state with infinite temperature, the state does not evolve and so trivially the power of the QB vanishes. However, with the charging being the \(\mathcal{PT}\)-symmetric Hamiltonian, the process is no longer unitary and it is debatable whether we can extract power even at high temperature as seen from Fig. \ref{fig:power_thermal}. The non-monotonic behavior of \(P_{\max}\) with respect to \(\beta\) cannot be observed, when the charging Hamiltonian is Hermitian, thereby emphasizing that our selection of a non-Hermitian charging Hamiltonian introduces a nontrivial aspect to our paper.

\begin{figure}
    \centering
    \includegraphics[scale=0.5]{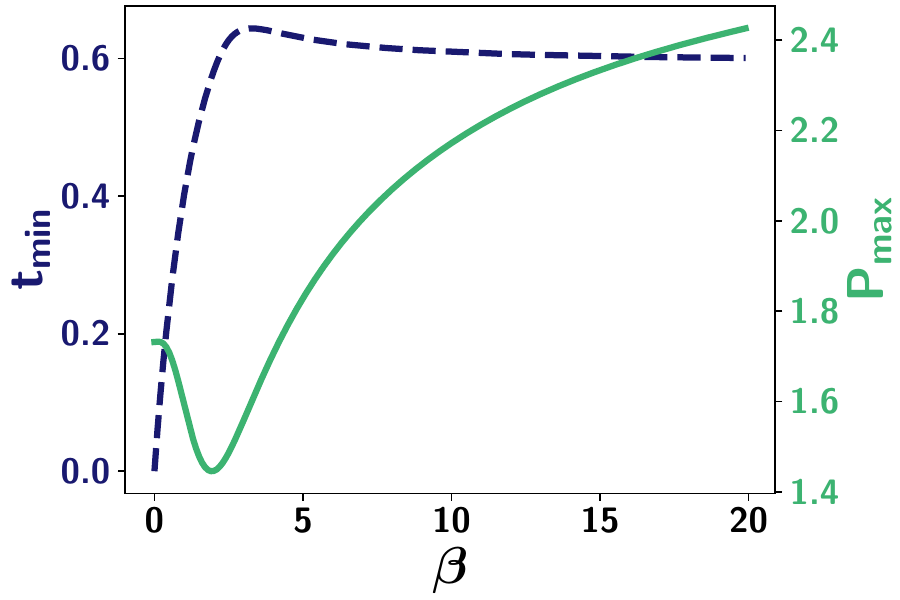}
    \caption{(Color online) \textbf{Temperature-dependence of \(P_{\max}\) and minimum time taken to achieve \(P_{\max}\), denoted by \(t_{\min}\). }
   \(P_{\max}\) (solid line (right ordinate)) and \(t_{\min}\) (dashed line (left ordinate)) vs. $\beta = 1/k_BT$ (abscissa) for  \(\alpha_i =2 \pi/3\).
   The initial state of the QB is prepared as the canonical equilibrium state of the $XX$ model while the charger is the \(\mathcal{PT}\)-symmetric one with \(\alpha_i = 2 \pi/3\). Notice that the decreasing behavior of the \(P_{\max}\) as well as \(t_{\min}\) with the increase of temperature is same as typically observed in the Hermitian domain. However, close to high temperature, in the framework of the non-Hermitian systems, nonmonotonic behaviors for both the quantities emerge with \(\beta\). 
   All the axes are dimensionless.}
    \label{fig:power_time_temp}
\end{figure}

 Towards explaining this nonmonotonicity, we compute the minimum time taken to reach the maximum power, \(t_{\min}\) during charging. It is motivated by a recent work   \cite{faster_bender_2007} in which it was shown that non-Hermitian evolution occurs at a faster pace compared to its Hermitian counterparts. 
 We observe that \(t_{\min}\) also exhibits a non-monotonic behavior for small values of \(\beta\) in the presence of strong non-Hermiticity parameter \(\alpha_i\) in the charging Hamiltonian as shown in  Fig. \ref{fig:power_time_temp} (solid (green) line). 
We can argue that along with other system parameters in the battery Hamiltonian, the nonmonotonic nature of \(t_{\min}\) with \(\beta\) also attributes to the non-monotonicity in \(P_{\max}\). 



\section{Charging battery with $\mathcal{RT}$ symmetric Hamiltonian}
\label{sec:RTsymm}

Let us reverse the design of the QB and check whether the benefit due to non-Hermiticity still persists or not. 
Instead of an interacting Hamiltonian as the QB, let us take the initial state as the ground state of the non-interacting Hamiltonian, given by 
\begin{equation}
    H_B^{n-int} = \sum_{j=1}^N\sigma^x_j.
    \label{eq:batterynonint}
\end{equation}
After normalizing the Hamiltonian, the eigenvector corresponding to the eigenvalue $-1$ is the initial state of the QB. 
A charging Hamiltonian in this case is taken to be the global non-Hermitian Hamiltonian, an $XY$ model with imaginary anisotropy parameter, having $\mathcal{RT}$ symmetry,  with  open boundary condition, represented as  
\begin{eqnarray}
        H_{charging}^\mathcal{RT}&=&\frac{J}{4}\sum_{j=1}^{N-1}\left[(1+i\gamma)\sigma^{x}_{j}\sigma^{x}_{j+1}+(1-i\gamma)\sigma^{y}_{j}\sigma^{y}_{j+1}\right]\nonumber\\&&+\frac{h'}{2}\sum_{j=1}^{N}\sigma_j^z,
        \label{eq:chargerRT}
    \end{eqnarray}
    where the operator $\mathcal{R}$  rotates the spin by $\frac{\pi}{2}$, i.e.,  \(\mathcal{R} \equiv e ^{\left[-i(\pi / 4) \sum_{j=1}^{N} \sigma_{j}^{z}\right]}\) and  $\mathcal{T}$ is  again the complex conjugation.
    Note that the charging Hamiltonian does not individually commute with either of the operators, $[H_{charging}^\mathcal{RT}, \mathcal{R}] \ne 0$ or $[H_{charging}^\mathcal{RT}, \mathcal{T}] \ne 0$ although 
$[H_{charging}^\mathcal{RT}, \mathcal{RT}] = 0$, thereby making it a pseudo-Hermitian Hamiltonian. It has been shown that in the symmetry unbroken phase, the Hamiltonian has real eigenvalues while it contains complex conjugated imaginary eigenvalues in the broken phase \cite{SongReal} and the transition occurs when \(h \equiv h'/|J| = \sqrt{1 + \gamma^2}\). 
Notice that in this scenario, when the initial battery Hamiltonian is non-interacting, the interacting Hamiltonian is necessary to charge the battery for obtaining the quantum advantage (quadratic scaling of power) which cannot be generated by the non-interacting charger \cite{globalcharging}.
   When the charging Hamiltonian is  Hermitian, \(i\gamma\)  is replaced by \(\gamma\) and is denoted by \(H_{charging}^{Herm}\). Moreover, the magnetic fields of non-Hermitian and Hermitian charging Hamiltonian, \(H_{charging}^{\mathcal{RT}}\) and \(H_{charging}^{Herm}\),  are denoted as \(h_i\) and \(h_r\) respectively.

    This type of Hamiltonian in Eq. (\ref{eq:chargerRT}) can be realized when a spin chain with \(XX\)-interaction is influenced by an environment.   A dissipative coupling between neighboring sites can be introduced through reservoir engineering, which has recently been explored, revealing nonreciprocal photon transmission, persistent currents, and other intriguing phenomena \cite{metelmann_prx_2015, keck_pra_2018,roccati_osi_2022}. The  evolution of the system, in this situation,  is governed by the GKLS master equation due to the presence of the environment as
\begin{equation}
    \frac{d\rho}{dt}=-[H_{S},\rho]+\kappa \sum_{j}\mathcal{L}[\sigma_j^-](\rho)+\sum_{j}\mathcal{L}[g_j(\{\sigma\})](\rho).
    \label{eq:master_eqn}
\end{equation} 
Here \(\mathcal{L}[A]\) represents the Lindblad operators associated with environmental effects such that the second term denotes the local dissipation, and the third term corresponds to the non-local dissipation between the sites. The Lindblad operator is defined as \(\mathcal{L}[A] = A\rho A^\dagger-\frac{1}{2}\{A^\dagger A,\rho\}\), and the expression for \(g_j(\{\sigma\})\) is given by 
 \begin{equation}
    g_j(\{\sigma\})=p \sigma_j^-+q \sigma_j^++r\sigma_{j+1}^-+s\sigma_{j+1}^+ ,
    \label{eq:lidblad}
\end{equation} 
where \(p\), \(q\), \(r\), and \(s\) represent suitable coupling parameters with the correlated environment, which can be complex, in general. To produce an $\mathcal{RT}$-symmetric Hamiltonian, we set \(p\) and \(s\) to be zero, while \(q = -{\gamma}/{\sqrt{2}}\), and \(r = {\gamma}/{\sqrt{2}}\). Therefore, the resulting effective Hamiltonian can be written as
\begin{eqnarray}
    \nonumber H_{eff} &=& H_{S} - \frac{i}{2}\sum_{j}\mathcal{L}[g_j(\{\sigma\})]^\dagger \mathcal{L}[g_j(\{\sigma\})]\\
    \nonumber &=& H_{S} - \frac{i\gamma}{4}\sum_{j}(\sigma_j^+-\sigma_{j+1}^-)(\sigma_j^--\sigma_{j+1}^+)\\ 
    &=& \sum_{j} (h+\frac{i\gamma}{4})\sigma_j^z + \frac{1}{4}\Big ((1+i\gamma)\sigma_j^x\sigma_{j+1}^x\\
    \nonumber &&+(1-i\gamma)\sigma_j^y\sigma_{j+1}^y\Big).
\end{eqnarray}
In this Hamiltonian, there are two dissipative terms -- one represents the dissipative coupling between subsystems, and the other one is a local dissipative term represented as \((h+\frac{i\gamma}{4})\)
which can independently be modified by tuning the local dissipative environment. 
By neglecting the local dissipative terms,  the effective Hamiltonian reduces to \(H_{iXY}\), given in Eq. (\ref{eq:chargerRT}).


 We will first demonstrate that the non-Hermitian charging Hamiltonian can produce more power than its Hermitian counterparts from a QB having two lattice sites.

   

\begin{proposition}
    The maximum power stored (\(P^{\mathcal{RT}}_{\max}\)) of a two-site quantum battery due to the $\mathcal{RT}$-symmetric $XY$ charger with transverse magnetic field is greater than that of the Hermitian $XY$ model provided the strength of the applied magnetic field is small and is strictly less than unity.
\end{proposition}

\begin{figure}[h]
    \centering
    \includegraphics[width=\linewidth]{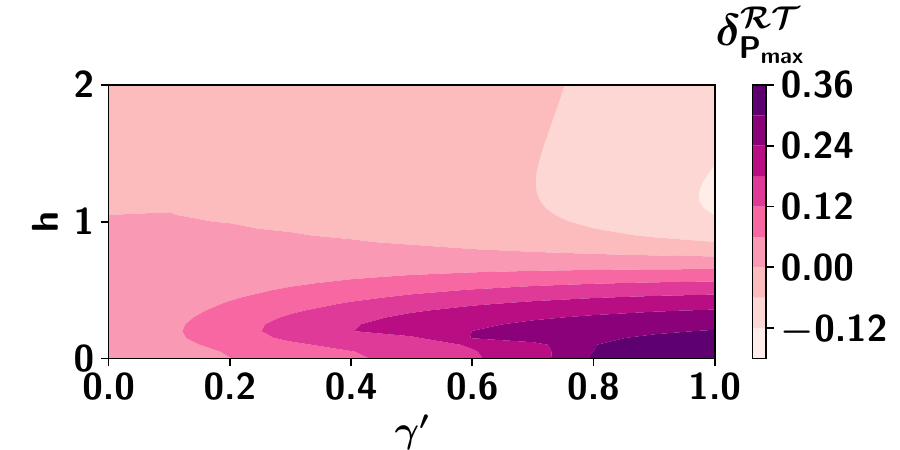}
    \caption{\textbf{Non-Hermitian effects on the QB.} Map plot of
    \(\delta_{P_{\max}}^{\mathcal{RT}}\) with the variation of parameters in the charging Hamiltonian, \(\gamma^\prime\) (horizontal axis) and \(h\) (vertical axis). The initial state is the ground state of the non-interacting battery Hamiltonian, \(H_B^{n-int}\) given in \textcolor{blue}{Eq. (\ref{eq:batterynonint})}.  Note that in Fig. \ref{fig:power_nh_h}, the difference was plotted with respect to the battery Hamiltonian. However, both the plots manifest some advantage in the presence of the non-Hermitian charger over the Hermitian one.  
    Here $N=2$. Both the axes are dimensionless.  }
    \label{fig:g_h_n2}
\end{figure}

\begin{proof} To prove it, we compare the cases when the charging is Hermitian, i.e., $\gamma = -i\gamma^\prime$ and when it is non-Hermitian, $\gamma = \gamma^\prime$. 
We take the ground state of the normalized Hamiltonian, \(H^{n-int}_B\), $|\psi(0)\rangle=\frac{1}{2} \left(
\begin{array}{c}
 1 \\
 -1 \\
 -1 \\
 1 \\
\end{array}
\right),$ 
with eigenvalue $ = -1$ as the initial state of the QB. After applying the evolution due to the charging, the evolved state is, $|\psi(t)\rangle = \frac{e^{-iH_{charging}^\mathcal{RT}t}|\psi(0)\rangle}{\mbox{tr}[e^{-iH_{charging}^\mathcal{RT}t}|\psi(0)\rangle]} = |\psi(t)\rangle = \frac{1}{\sqrt{\mathcal{N}}} \left(
\begin{array}{c} A \\ B \\ B \\ C \end{array} \right )$, where the expressions for \(A, B\), \(C\) and \(\mathcal{N}\) are given in Appendix. It is possible to compute \(P^{\mathcal{RT}} (t)\) and the corresponding \(P^{Herm}(t) \) (see Appendix) and hence again we compute the difference, given by
\begin{equation}
    \delta^{\mathcal{RT}}_{P_{\max}} = \max_t (P^{\mathcal{RT}}(t))-\max_t(P^{Herm}(t)) = P_{\max}^\mathcal{RT}-P_{\max}^{\text{Herm}}
    \label{eq:diffRT}
\end{equation}
for $\gamma^\prime \in \{0,1\}$ and $h \in \{0,2\}$. 
As shown in Fig. \ref{fig:g_h_n2}, there exists a region of \(h\), i.e., when \(h <0.8\), \(\delta^{\mathcal{RT}}_{P_{\max}} >0\). It implies that the non-Hermitian charger clearly gives some benefit over the Hermitian ones.  
\end{proof}

\emph{Remark.} The upper bound on \(h\) shown in Proposition 2 which is not unity is possibly due to the finite size effect. We will also show that with a moderate system size, the battery with a non-Hermitian charger provides a higher maximal power than that of the Hermitian ones when \(h <1.0\), irrespective of the values of \(\gamma\) which controls its non-Hermiticity in the later part of the section (see Fig. \ref{fig:igamma_gamma_n}) (we
drop the superscripts $\mathcal{RT}$ and ``Herm" in the analysis as it will
be evident from the context). 



\subsection{Effect of \(\mathcal{RT}\)-symmetric charger on power}

\begin{figure}[h]
    \centering
    \includegraphics[width=\linewidth]{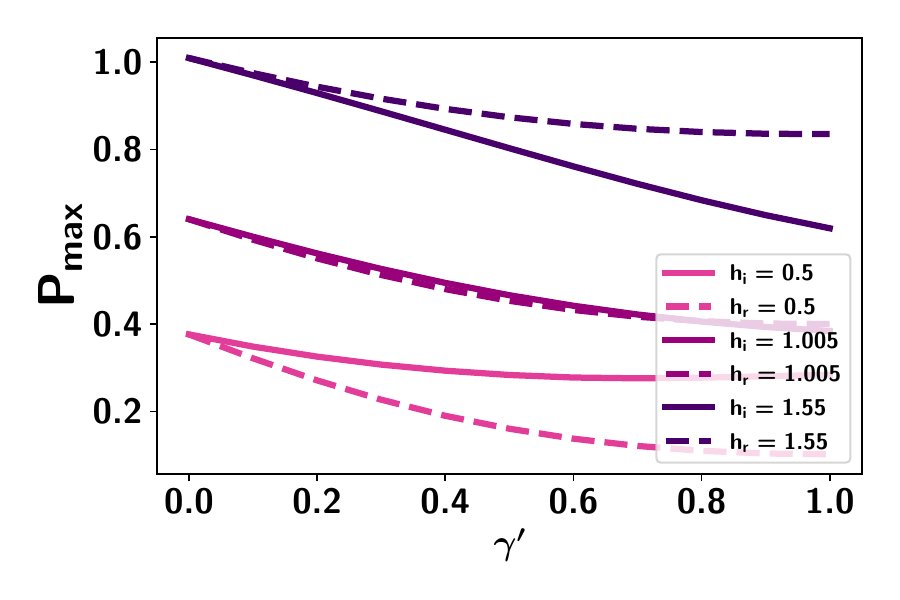}
    \caption{{\bf Comparison between Hermitian and non-Hermitian chargers.} \(P_{\max}\) (ordinate) vs. \(\gamma'\) (abscissa). Solid lines represent non-Hermitian charger in Eq. (\ref{eq:chargerRT}) (\(\gamma = \gamma'\)) while dashed lines represent the Hermitian ones (\(\gamma = - i \gamma'\)), representing the \(XY\) model. The ground state of the non-interacting battery Hamiltonian represents the initial state of the QB as in Fig. \ref{fig:g_h_n2}. Here $h_i$ and $h_r$ represent the non-Hermitian and  Hermitian chargers respectively.
    The system size is taken to be eight, i.e., \textcolor{blue}{N=8}.  Both the axes are dimensionless. } 
    \label{fig:rt_h_nh}
\end{figure}

We compare the maximum power generated via the non-Hermitian model corresponding to the applied magnetic field, denoted with $h_i$ and  \(P_{\max}\) produced by the Hermitian model having applied field $h_r$ with the variation of \(\gamma'\) in Fig. \ref{fig:rt_h_nh}.  It is evident that the difference between generated power by non-Hermitian and Hermitian chargers, \(\delta_{P_{\max}}^{\mathcal{RT}}\), is maximum when \(h_{i(r)}\) is small,   decreases with the increase of \(h_{i(r)}\), and finally becomes negative for high values of \(h_{i(r)}\), i.e., when \(h_{i(r)} >1\). More precisely, the observations can be listed as follows.   


\begin{enumerate}
    \item  When $h_{i(r)} \sim 0.5 < 1.0$, the non-Hermitian charging admits higher $P_{max}$  compared to the Hermitian ones  $\forall \text{  } \gamma^\prime$. For high \(\gamma'\), eg. for $\gamma^\prime \ge 0.6$, the maximum generated power, $P_{max}$, is small for the Hermitian case and in this regime, the non-Hermitian advantage is more pronounced than that of low \(\gamma'\).  
     
     \item  Let us consider the case with $h_{i(r)} \sim 1.005$.  In this domain, both non-Hermitian and Hermitian charging lead to almost the same $P_{max}$ value, thereby exhibiting no advantage. Interestingly, \(\delta_{P_{\max}}^{\mathcal{RT}}\) vanishes with the increase of \(N\). 
      \item  Going beyond $h_{i(r)} >1$, eg.,  $1.5$, the performance of the QB in terms of \(P_{\max}\) with the Hermitian charger  outperforms the corresponding non-Hermitian QB.  
\end{enumerate}
Therefore, close inspection reveals that like the \(\mathcal{PT}\)-symmetric charger, the \(\mathcal{RT}\)-symmetric charging Hamiltonian has the potential to give benefit provided the charging Hamiltonian is tuned in a suitable way. Note that the initial battery Hamiltonian also plays an important role in obtaining gain from the non-Hermitian charger. For example, if one chooses a battery Hamiltonian to be a nearest-neighbor Ising model, \( H_B= \sum_{i = 1}^N \sigma_i^x \sigma_{i+1}^x\) and the same non-Hermitian charger in Eq. (\ref{eq:chargerRT}) is used to evolve the ground state of such an interacting Hamiltonian, there exists a region of parameters for which advantage is absent (\(\delta_{P_{\max}}^{\mathcal{RT}}<0\)) for \(N\ge 2\) while the noninteracting battery Hamiltonian in Eq. (\ref{eq:batterynonint}) performs better \((\delta_{P_{\max}}^{\mathcal{RT}}>0\)) for the same set of parameters.

\subsection{Effect of system size on power} 
\begin{figure}[h]
    \centering
    \includegraphics[width=\linewidth]{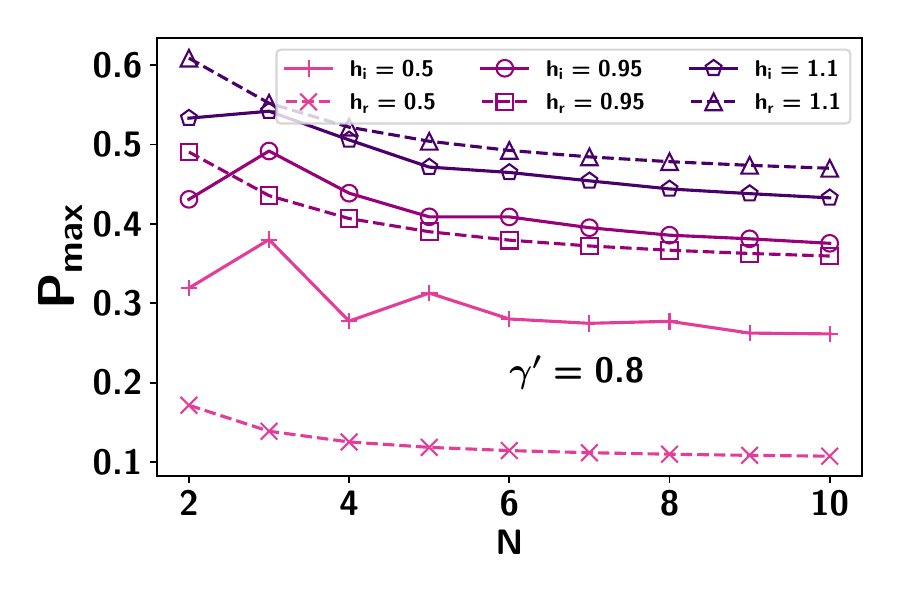}
    \caption{\textbf{Scaling of the QB with \(\mathcal{RT}\)-symmetric charger:} $P_{\max}$ (\(y\)-axis) vs. $N$ (\(x\)-axis) for $\gamma'=0.8$. 
    All other specifications are the same as in Fig. \ref{fig:rt_h_nh}. Both the axes are dimensionless. 
    }
    \label{fig:igamma_gamma_n}
\end{figure}

It is natural to ask whether the improvements remain valid even when one wants to design a battery with a reasonable system size.  Until now, it has been exhibited for 
$N=2 \text{ and } 6$. For a fixed \(\gamma'\),   we check whether the advantage is just a numerical artifact or not by comparing $P_{max}$ with $N$ for different exemplary values of $h_{i(r)}$. As depicted in Fig. \ref{fig:igamma_gamma_n}, we observe that \(P_{\max}\) saturates to a fixed value with the system-size \(N\) irrespective of  \(\gamma'\) values and the strength of the magnetic fields, \(h_i\) and \(h_r\).  Hence \(P_{\max}\) of a QB consisting of a reasonable number of lattice sites continues to be advantageous in the non-Hermitian case provided the magnetic field in the charger is adjusted properly.


 
 \subsection{ Temperature dependence of power.} 
\begin{figure}[h]
      \centering
      \includegraphics[width=\linewidth]{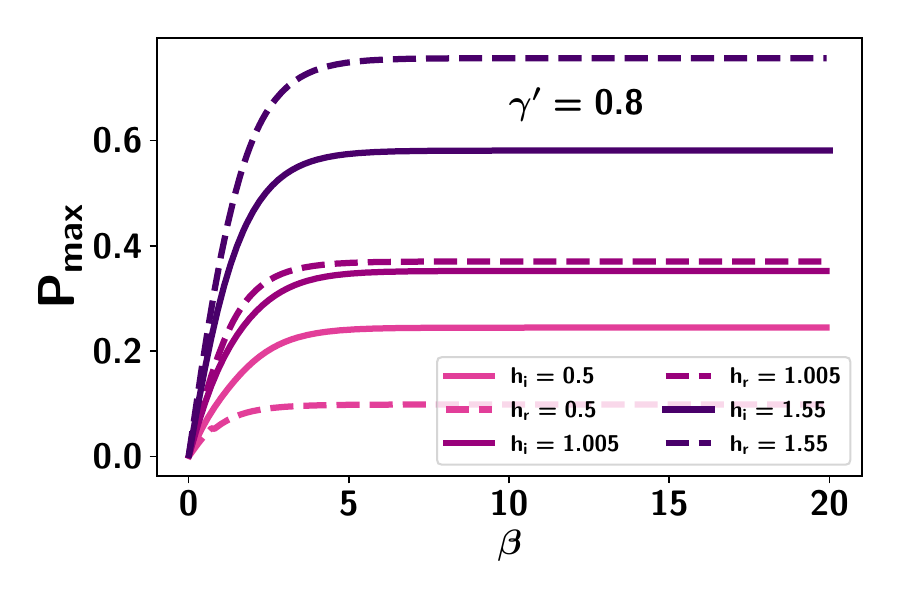}
      \caption{{\bf Effects of thermal fluctuations on the non-Hermitian battery:} The maximal power (ordinate) against \(\beta = \beta'/|J|\) (abscissa) where the initial state is prepared as the thermal state of the battery. The charging is again by the \(\mathcal{RT}\)-symmetric Hamiltonian with \(\gamma'= 0.8\). All other specifications are the same as in Fig. \ref{fig:rt_h_nh}. Both the axes are dimensionless.   
      }
      \label{fig:thermal_RT}
  \end{figure}

We have already observed some non-trivial effects on the power output of the QB with \(\mathcal{PT}\)-symmetric charger when the initial state is the thermal state, \(\rho_{\beta}\) with \(\beta = \beta'/|J|\). It increases with the decrease of temperature (see Fig. \ref{fig:thermal_RT}), thereby showing detrimental effects on power in the presence of thermal fluctuation. Like the \(\mathcal{PT}\)-symmetric case, \(P_{\max}\) is close to zero in the limiting case, i.e., \(\beta \rightarrow 0\) although it does not vanish exactly like the unitary dynamics.  Interestingly, however, \(\delta_{P_{\max}}^{\mathcal{RT}}\) is small when the temperature is moderately high. In other words, the superiority of non-Hermitian (Hermitian)  systems over Hermitian (non-Hermitian) ones gets pronounced with a moderate temperature of the initial state of the QB. 
  
  
\section{Conclusion}
\label{sec:conclu}

The dynamics of quantum systems governed by the non-Hermitian Hamiltonian have attracted lots of attention in recent times. On the other hand, the evolution of a quantum system plays an important role in building quantum technologies, like thermal machines. Among several quantum thermal devices, a prominent one is the quantum battery which shows a better storage capacity with the help of quantum mechanics than the classical models.


We incorporated non-Hermitian evolution in constructing quantum QBs. Specifically, we used both \(\mathcal{PT}\)- and \(\mathcal{RT}\)-symmetric charging Hamiltonian to charge the ground state of an interacting and non-interacting
Hamiltonian respectively. We also provided possible realizations of such chargers as an effective description of quantum systems interacting with an environment. 
When the battery consists of two sites,  we analytically proved that the maximum power with non-Hermitian chargers gets enhanced compared to their Hermitian counterparts provided the system parameters are tuned appropriately.  
In the case of a local \(\mathcal{PT}\)-symmetric charger, when the initial state of the QB is the ground state of the $XY$ model with the transverse magnetic field having a moderate system size, we demonstrate that it can produce extractable power which cannot be obtained with the QB Hamiltonian without interactions,  thereby showing quantum advantage. Moreover, we find that the power scales with  \(\sqrt{\text{system size}}\), thereby exhibiting the persistence of non-Hermitian advantage even in the macroscopic limit.

Starting with the ground state of the non-interacting Hamiltonian, we demonstrated that the interacting \(\mathcal{RT}\)-symmetric charger has also the potential to generate a higher amount of power in the QB than that of the corresponding Hermitian charger provided the magnetic field in the charging is adjusted appropriately. We also observed that the power also saturates to a nonvanishing finite value both in the Hermitian and non-Hermitian scenarios with the increase of system size. 

Beyond the zero-temperature scenario, if the initial state of the battery is the thermal state and the charging Hamiltonian is non-Hermitian, interesting non-trivial results emerge -- as expected, the maximum power decreases with the increase of temperature although unlike Hermitian systems, it does not vanish at infinite temperature. 

The construction of a quantum battery proposed in the framework of non-Hermitian quantum mechanics and the advantages reported here open up a possibility to design other quantum technologies including quantum heat engines, and refrigerators in this paradigm.  It will be an interesting direction to explore the possible implementations of these devices using currently available technologies.


\acknowledgements

We acknowledge the support from the Interdisciplinary Cyber Physical Systems (ICPS) program of the Department of Science and Technology (DST), India, Grant No.: DST/ICPS/QuST/Theme- 1/2019/23, the use of \href{https://github.com/titaschanda/QIClib}{QIClib} -- a modern C++ library for general purpose quantum information processing and quantum computing (\url{https://titaschanda.github.io/QIClib}), and the cluster computing facility at the Harish-Chandra Research Institute. This research was supported in part
by the ’INFOSYS scholarship for senior students.

\appendix

\section{Analytical expression of power in $\mathcal{PT}$- and $\mathcal{RT}$- symmetric charging }
\label{sec:appendix}
\begin{widetext}
\subsection{Battery based on \(\mathcal{PT}\)-symmetric charger} The evolved state of a system consisting of two lattice sites at later time $t$ with the evolution operator constructed via \(H^{\mathcal{PT}}_{charging}\) reads as 

\begin{equation}
\ket{\psi(t)}=\left(
\begin{array}{c}
 -\frac{ \cos ^2 \alpha_i  \csc ^2 (t \cos \alpha_i )}{\cos ^4(\alpha_i +t \cos \alpha_i ) \csc ^4(t \cos \alpha_i )+\cos ^2(\alpha_i +t \cos \alpha_i ) \csc ^2(t \cos \alpha_i )+1} \\
 -\frac{i\cos ^2(\alpha_i ) \cos (\alpha_i +t \cos (\alpha_i )) \sin (t \cos (\alpha_i ))}{\cos ^4(\alpha_i +t \cos (\alpha_i ))+2\sin ^2(t \cos (\alpha_i )) \cos ^2(\alpha_i +t \cos (\alpha_i ))+\sin ^4(t \cos (\alpha_i ))} \\
 -\frac{i\cos ^2(\alpha_i ) \cos (\alpha_i +t \cos (\alpha_i )) \sin (t \cos (\alpha_i ))}{\cos ^4(\alpha_i +t \cos (\alpha_i ))+2\sin ^2(t \cos (\alpha_i )) \cos ^2(\alpha_i +t \cos (\alpha_i ))+ \sin ^4(t \cos (\alpha_i ))} \\
 \frac{\cos ^2(\alpha_i )}{ \sec ^2(\alpha_i +t \cos (\alpha_i )) \sin ^4(t \cos (\alpha_i ))+\sin ^2(t \cos (\alpha_i ))+\cos ^2(\alpha_i +t \cos (\alpha_i ))} \\
\end{array}
\right),
\end{equation}
where the initial state $\ket{\psi(0)}=\ket{0001}$ is the ground state of the Hamiltonian, \(H_B\) when \(J \in [-2h, 2h -0.1]\).
We can  express the form of the power which depends on the parameter of the system, given by
\begin{equation}
      P^\mathcal{PT}(t,h,J,\alpha_i) = \frac{-h \cos ^4(\alpha_i +t \cos (\alpha_i ))+h \sin ^4(t \cos (\alpha_i ))+J \cos ^2(\alpha_i +t \cos (\alpha_i )) \sin ^2(t \cos (\alpha_i ))}{h t \left(\cos ^4(\alpha_i +t \cos (\alpha_i ))+2\cos ^2(\alpha_i +t \cos (\alpha_i )) \sin ^2(t \cos (\alpha_i ))+\sin ^4(t \cos (\alpha_i ))\right)}+\frac{1}{t}.
    \label{eq:power_nherm}
\end{equation}
We calculate the power generated when the $H_{charging}^\mathcal{PT}$ is replaced with its Hermitian counterpart as 
 \begin{equation}
    H_{charging}^{Herm}=\sum_{j=1}^{N}\sigma_j^x+\sin\alpha_r\sigma_j^z.
\end{equation}
In the Hermitian domain, the average power takes the form as 
\begin{equation}
    P^{Herm}(t,h,J,\alpha_r)=\scriptsize	{\frac{-h \cos 4 \alpha_r +\cos 2 \alpha_r  (8h-2J)+\cos \left(t \sqrt{6-2 \cos 2 \alpha_r }\right) (\cos 2 \alpha_r  (4h+2J)-12h-2J)-7h-J \cos \left(2 t \sqrt{6-2 \cos 2 \alpha_r }\right)+3J}{h t (-6 \cos 2 \alpha_r +0.5 \cos 4 \alpha_r +9.5)}+\frac{1}{t}}. 
    \label{eq:power_herm}
\end{equation}
Comparing \(P^\mathcal{PT}(t,h,J,\alpha_i)\) and \( P^{Herm}(t,h,J,\alpha_r)\), and optimizing over time, we can find that the difference, \(\delta_{P_{\max}}^{\mathcal{PT}}
\) is positive in \(J \in [-2h, 2h -0.1]\), thereby establishing non-Hermitian enhancement.

\subsection{QB with \(\mathcal{RT}\)-symmetric charger} The ground state of the normalized Hamiltonian, \(H^{n-int}_B\) which is the initial state of the QB reads as
\begin{eqnarray}
    |\psi(0)\rangle=\frac{1}{2} \left(
\begin{array}{c}
 1 \\
 -1 \\
 -1 \\
 1 \\
\end{array}
\right),
\end{eqnarray}
with eigenvalue $ = -1$. The evolution operator based on \(H^{\mathcal{RT}}_{charging}\) acts on the initial state and produces the  evolved state, given by 
\begin{eqnarray}
|\psi(t)\rangle =
\frac{1}{\mbox{tr}(e^{-iH_{charging}^\mathcal{RT}t}|\psi(0)\rangle )} e^{-iH_{charging}^\mathcal{RT}t}|\psi(0)\rangle,  
\end{eqnarray}
which reduces to  
\begin{eqnarray}
    |\psi(t)\rangle = \frac{1}{\sqrt{\mathcal{N}}} \left(
\begin{array}{c} A \\ B \\ B \\ C \end{array} \right ). 
\end{eqnarray}


Here 
\begin{eqnarray}
&& A = \frac{\gamma \sinh \left(\frac{1}{2} t \sqrt{\gamma^2-4 h^2}\right)}{2 \sqrt{\gamma^2-4 h^2}}+\frac{1}{2} \left(\cosh \left(\frac{1}{2} t \sqrt{\gamma^2-4 h^2}\right)-\frac{2 i h \sinh \left(\frac{1}{2} t \sqrt{\gamma^2-4 h^2}\right)}{\sqrt{\gamma^2-4 h^2}}\right), \nonumber\\
&& B = -\frac{\cos(t/2)}{2}  + \frac{i\sin(t/2)}{2}, \nonumber\\
&& C = \frac{\gamma \sinh \left(\frac{1}{2} t \sqrt{\gamma^2-4 h^2}\right)}{2 \sqrt{\gamma^2-4 h^2}}+\frac{1}{2} \left(\cosh \left(\frac{1}{2} t \sqrt{\gamma^2-4 h^2}\right)+\frac{2 i h \sinh \left(\frac{1}{2} t \sqrt{\gamma^2-4 h^2}\right)}{\sqrt{\gamma^2-4 h^2}}\right), \nonumber \\
&& \text{and}\,\, \mathcal{N} = \frac{\gamma \sqrt{\gamma^2-4 h^2} \sinh \left(t \sqrt{\gamma^2-4 h^2}\right)+\gamma^2 \cosh \left(t \sqrt{\gamma^2-4 h^2}\right)+\gamma^2-8 h^2}{2 \gamma^2-8 h^2} \nonumber.
\end{eqnarray}
We compute the power in Eq. (\ref{Eq:maxP}), when the charging is performed by  $H_{charging}^{\mathcal{RT}}$ with $\gamma = \gamma^\prime$, given by
\begin{equation}
   P^{\mathcal{RT}}(t) =  \frac{2 \cos \left(\frac{t}{2}\right) \left(\left({\gamma^{\prime}}^2-4 h^2\right) \cos \left(\frac{t}{2}  \sqrt{4 h^2-{\gamma^{\prime}}^2}\right)-{\gamma^{\prime}} \sqrt{4 h^2-{\gamma^{\prime}}^2} \sin \left(\frac{t}{2}  \sqrt{4 h^2-{\gamma^{\prime}}^2}\right)\right)}{t \left| \cos \left(\sqrt{4 h^2-{\gamma^{\prime}}^2} t\right) {\gamma^{\prime}}^2+{\gamma^{\prime}}^2-\sqrt{4 h^2-{\gamma^{\prime}}^2} \sin \left(\sqrt{4 h^2-{\gamma^{\prime}}^2} t\right) g-8 h^2\right| }+\frac{1}{t}, \text{ when } {\gamma^{\prime}}^2 < 4h^2, 
\end{equation}
and 
\begin{equation}
   P^{\mathcal{RT}}(t)=  1-\frac{2 \cos \left(\frac{t}{2}\right) \left({\gamma^{\prime}} \sqrt{{\gamma^{\prime}}^2-4 h^2} \sinh \left(\frac{1}{2} t \sqrt{{\gamma^{\prime}}^2-4 h^2}\right)+\left({\gamma^{\prime}}^2-4 h^2\right) \cosh \left(\frac{1}{2} t \sqrt{{\gamma^{\prime}}^2-4 h^2}\right)\right)}{t \left| \cosh \left(\sqrt{{\gamma^{\prime}}^2-4 h^2} t\right) {\gamma^{\prime}}^2+{\gamma^{\prime}}^2+\sqrt{{\gamma^{\prime}}^2-4 h^2} \sinh \left(\sqrt{{\gamma^{\prime}}^2-4 h^2} t\right) {\gamma^{\prime}}-8 h^2\right| }, \text{ when } {\gamma^{\prime}}^2 > 4h^2.
\end{equation}
When the charger is  Hermitian,  $H_{charging}^{Herm}$ with $\gamma = -i\gamma^\prime$, 
the generated power can be computed as
\begin{equation}
    P_{Herm}(t) = 1-\frac{\frac{{\gamma^{\prime}} \sin \left(\frac{t}{2}\right) \sin \left(\frac{1}{2} t \sqrt{{\gamma^{\prime}}^2+4 h^2}\right)}{\sqrt{{\gamma^{\prime}}^2+4 h^2}}+\cos \left(\frac{t}{2}\right) \cos \left(\frac{1}{2} t \sqrt{{\gamma^{\prime}}^2+4 h^2}\right)}{t}, \text{ when } {\gamma^{\prime}}^2 + 4h^2 \ne 0.
\end{equation}

To demonstrate the benefit of non-Hermitian charger, we introduce a quantity, \(\delta^{\mathcal{PT}(\mathcal{RT})}_{\text{max}}\), which represents the difference between the maximum power achieved through non-Hermitian and Hermitian charging processes, as given in Eqs. (\ref{eq:diffPT}) and (\ref{eq:diffRT}). A positive value of \(\delta^{\mathcal{PT}(\mathcal{RT})}_{\text{max}}\) signifies the advantageous role played by non-Hermiticity. Consequently, our paper focuses on identifying regions within the parameter space where \(\delta^{\mathcal{PT}(\mathcal{RT})}_{\text{max}} > 0\), and we highlight regions in Figs  \ref{fig:power_nh_h} and \ref{fig:g_h_n2}.  Our paper reveals that the presence of \(\mathcal{PT}\) (\(\mathcal{RT}\))-symmetry within the charging Hamiltonian yields favorable effects on energy storage within the quantum battery, surpassing those of the Hermitian counterpart.

\end{widetext}

\bibliography{ref.bib}

\end{document}